\documentclass[twocolumn]{IEEEtran}
\usepackage{amsmath}
\usepackage{amssymb}
\usepackage{amsfonts}
\usepackage{graphicx}
\usepackage{epsfig}
\usepackage{subfigure}
\usepackage{psfrag}
\usepackage{cite}
\usepackage{latexsym}
\usepackage{url}
\usepackage{color}
\usepackage{bm}
\usepackage{algorithm}

\begin{document}
\title{Time-Division Energy Beamforming for Multiuser Wireless Power Transfer with Non-Linear Energy Harvesting
\thanks{G. Ma is with the School of Information Engineering, Guangdong University of Technology, Guangzhou, China, and the Future Network of Intelligence Institute (FNii), The Chinese University of Hong Kong (Shenzhen), Shenzhen, China (e-mail: gangma.gdut@gmail.com). }
\thanks{J. Xu is with the Future Network of Intelligence Institute (FNii) and the School of Science and Engineering, The Chinese University of Hong Kong (Shenzhen), Shenzhen, China (e-mail: xujie@cuhk.edu.cn). J. Xu is the corresponding author.}
\thanks{Y.-F. Liu is with the State Key Laboratory of Scientific and Engineering Computing, Institute of Computational Mathematics and Scientific/Engineering Computing, Academy of Mathematics and Systems Science, Chinese Academy of Sciences, Beijing, China (e-mail: yafliu@lsec.cc.ac.cn).}
\thanks{M. R. V. Moghadam is with the TransferFi Pte Ltd, Singapore (e-mail: reza@transferfi.com).}}
\author{Ganggang Ma, Jie Xu, Ya-Feng Liu, and Mohammad R. Vedady Moghadam \vspace{-1.5em}}
\maketitle

\begin{abstract}
Energy beamforming has emerged as a promising technique for enhancing the energy transfer efficiency of wireless power transfer (WPT). However, the performance of conventional energy beamforming may seriously degrade due to the non-linear radio frequency (RF) to direct current (DC) conversion at energy receivers (ERs). To tackle this issue, this letter proposes a new time-division energy beamforming, in which different energy beamforming matrices (of high ranks in general) are time shared to exploit the ``convex-concave'' shape of the RF-DC power relation at ERs. By considering a particular time duration for WPT, we maximize the minimum harvested DC energy among all ERs, by jointly optimizing the energy beamforming matrices and the corresponding time allocation. In order to solve the non-convex min-DC-energy maximization problem, we propose an efficient solution by using the techniques of alternating optimization and successive convex approximation (SCA). Numerical results show that the proposed time-division energy beamforming design indeed outperforms the conventional multi-beam and time-division-multiple-access (TDMA)-based energy transmissions.
\end{abstract}


\newtheorem{lemma}{\underline{Lemma}}[section]
\newtheorem{proposition}{\underline{Proposition}}[section]
\newtheorem{remark}{\underline{Remark}}[section]
\newtheorem{example}{\underline{Example}}[section]
\newcommand{\mv}[1]{\mbox{\boldmath{$ #1 $}}}

\section{Introduction}
Multi-antenna energy beamforming has been recognized as a promising technique to improve the end-to-end energy transfer efficiency for radio-signals-based wireless power transfer (WPT) \cite{Yong,Clerckx}. By deploying multiple antennas, energy transmitters (ETs) can properly adjust the transmit beamforming to steer wireless energy towards desirable directions, thus combating against the severe radio signal propagation loss, and charging intended energy receivers (ERs) more efficiently.

In the literature, there have been various prior works investigating the transmit energy beamforming design for multiuser WPT \cite{XuWCL} and the applications in simultaneous wireless information and power transfer (SWIPT) \cite{XuSWIPT,LiuSWIPT}, wireless powered communication networks (WPCN) \cite{WPCN}, mobile edge computing (MEC) \cite{Feng}, and Internet of things (IoT) \cite{DongIn}. For the ease of analysis, these prior works normally considered linear energy harvesting (EH) models at ERs, i.e., the radio frequency (RF) to direct current (DC) conversion efficiency at each ER is assumed to be constant regardless of the received RF power. In this case, it was shown in \cite{XuWCL} that for a multiuser WPT system, the multi-beam energy transmission design is optimal for maximizing the minimum harvested RF (or equivalently DC) power among all the ERs.

In practice, however, the RF-to-DC energy conversion efficiency is highly non-linear. In the literature, there are generally two different non-linear EH models to characterize such non-linear behaviors. The first non-linear EH model is obtained based on the Taylor expansion of diode characteristics, which has been widely used to facilitate the transmit signal waveform design \cite{Bruno,Reza}. By contrast, the second non-linear EH model specifies the relation between the input RF and output DC power based on sigmoid functions, which is determined via curve fitting by using practical measurement results, and is normally used to facilitate the energy beamforming design under fixed signal waveforms \cite{non-linearModel,Ma}. Based on the sigmoid function-based EH model, it is observed that the RF-DC power relation generally follows a ``convex-concave'' shape. Specifically, when the input RF power is less than a threshold, the DC power can be approximated as a convex function with respect to the RF power; otherwise, it is concave. Under this non-linear EH model, the performance of conventional multi-beam energy transmission (e.g., in \cite{LiuSWIPT}) is not optimal for multiuser WPT any longer. The performance loss is pronounced when the received RF power at ERs falls into the convex regime. This thus motivates this work to investigate new energy beamforming designs by considering the non-linear EH models.

In this letter, we study a multiuser multiple-input single-output (MISO) WPT system consisting of one ET with $M$ transmit antennas and $K$ ERs each with one single antenna, by considering the sigmoid function-based non-linear EH model. By considering a particular time duration, we aim to maximize the minimum harvested DC energy among all ERs, by jointly optimizing the energy beamforming matrices over time. First, under the non-linear EH model, we use a simple example to show that the time-division-multiple-access (TDMA)-based energy beamforming design can outperform the multi-beam energy transmission that was shown in \cite{LiuSWIPT} to be optimal under linear EH models. Motivated by this observation, we propose a more general beamforming design approach to exploit both benefits of multi-beam and TDMA-based energy transmissions, under which different energy beamforming matrices (of high ranks in general) are time shared to exploit the ``convex-concave'' conversion of the RF-DC power relation at ERs. To solve the non-convex min-DC-energy maximization problem, we propose an efficient solution by using the techniques of alternating optimization and successive convex approximation (SCA). Numerical results show that the proposed time-division energy beamforming design indeed outperforms the conventional multi-beam and TDMA-based energy transmission designs.

{\it Notation:} Boldface letters refer to vectors (lower case) or matrices (upper case). For a matrix $\mv S$, ${\mv S}^H$ and ${\mv S}^T$ denote its conjugate transpose and transpose, respectively. $\mv I$ denotes an identity matrix. ${\rm tr}(\cdot)$ denotes the trace of a squared matrix. $\mathbb{E}(\cdot)$ denotes the statistical expectation. $ \nabla f(x)$ denotes the first-derivative of $f(x)$ with respect to $x$. $\mathbb C^{x\times y}$ denotes the space of $x\times y$ complex matrices.

\section{System Model}
\begin{figure}
\centering
 \epsfxsize=1\linewidth
    \includegraphics[width=8.5cm]{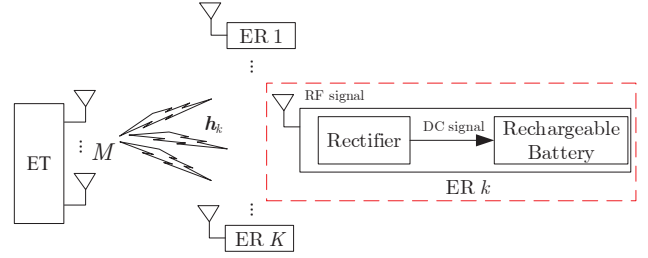}
\caption{The multiuser MISO WPT system.} \label{fig:Sys}
\vspace{-1em}
\end{figure}

As shown in Fig. \ref{fig:Sys}, we consider a multiuser MISO WPT system, where an ET transmits RF signals to wirelessly charge $K\ge1$ ERs. The ET is equipped with $M\ge1$ antennas and each ER is equipped with one single antenna. Let $\mathcal M \triangleq\{1,..., M\}$ and $\mathcal K \triangleq\{1,..., K\}$ denote the set of transmit antennas at the ET and the set of ERs, respectively.

We consider quasi-stationary channel models, in which the wireless channels remain unchanged within a particular time block of our interest. It is assumed that the ET knows the channel state information (CSI) from its antennas to all ERs. The CSI can be obtained by, e.g., using the energy feedback based channel estimation method \cite{EF}. Let $\mathcal T \triangleq (0,T]$ denote the time block with duration $T$. At any given time instant $t \in \mathcal T$, let $\mv s(t)\in\mathbb C^{M\times1}$ denote the energy-bearing signal transmitted from the ET, and $\mv h_k\in \mathbb C^{M\times1}$ denote the channel vector from the ET to the $k$-th ER. Then, the received RF power at the $k$-th ER is expressed as
\begin{align}
Q_k^{\rm RF}(t)=\mathbb{E}(|\mv h^{H}_k\mv s(t)|^2)=\mv h^H_k\mv S(t)\mv h_k,\ \forall k\in\mathcal K,
\label{equa:RF_1}
\end{align}
where $\mv S(t)\triangleq\mathbb{E}[\mv s(t)\mv s^H(t)]\in\mathbb C^{M\times M}$ denotes the transmit covariance matrix, which is positive semi-definite, i.e., ${\mv S}(t) \succeq \mv 0$.\footnote{Suppose that the rank of $\mv S(t)$ is given by $d \ge 1$ in general. This corresponds to the case when the ET transmits $d$ energy beamforming vectors at time $t$, each of which can be obtained via applying the eigenvalue decomposition on $\mv S(t)$.} Suppose that the maximum transmit power at the ET is $P_{\rm max}$, and then we have ${\rm tr}(\mv S(t))\le P_{\rm max}$.

As for the RF-to-DC energy conversion, we adopt the non-linear EH model in \cite{non-linearModel}, where the output DC power at the $k$-th ER is expressed as a sigmoid-like function with respect to the input RF power $Q_k^{\rm RF}(t)$, i.e.,
\begin{align}
Q^{\rm DC}_k(Q^{\rm RF}_k(t))
= \frac{Q^{\rm sig}_k(Q_k^{\rm RF}(t))-Q^{\rm max} \Omega}{1-\Omega},
\label{equa:non-linear}
\end{align}
with $\Omega=\frac{1}{1+e^{ab}}$ and $Q_k^{\rm sig}(Q_k^{\rm RF})= \frac{Q^{\rm max}}{1+e^{-a(Q_k^{\rm RF}-b)}}$. Here, $\Omega$ is a constant to ensure the zero-input/zero-output response for EH, $Q^{\rm max}$ denotes the maximum output DC power at the ERs when the rectifier is saturated, and $a$ and $b$ are two constants depending on the specific circuit. Based on (\ref{equa:non-linear}), it is clear that the relationship between the harvested DC power $Q_k^{\rm DC}(t)$ and the RF power $Q_k^{\rm RF}(t)$ has a convex-concave shape \cite{Clerckx,non-linearModel}. Accordingly, the average harvested DC energy over the whole block $\mathcal T$ at each ER $k$ is given by $\int_{0}^TQ^{\rm DC}_k(\mv h^H_k\mv S(t)\mv h_k){\rm d}t$.

For the considered multiuser MISO WPT setup, our objective is to maximize the minimum average harvested DC energy among all the ERs over time block $\mathcal T$ by optimizing the transmit energy covariance matrices ${\mv S}(t)$ over time. The optimization problem is formulated as
\begin{align}
{\mathtt{(P1)}}: \max\limits_{\bm{S}(t)}&~~\min\limits_{k\in\mathcal K}~\int_{0}^TQ^{\rm DC}_k(\mv h^H_k\mv S(t)\mv h_k){\rm d}t\label{equa:P0:1} \\
{\mathrm{s.t.}} &~~{\rm{tr}}({\mv{S}(t)})\le P_{\rm max},\ \forall t\in \mathcal T \label{equa:P0:max_power}\\
&~~\mv{S}(t) \succeq 0,\ \forall t\in \mathcal T. \label{equa:P0:tr_S}
\end{align}
Note that without loss of generality, in problem (P1) the transmit energy covariance matrices $\{\mv S(t)\}$ are adjustable over time. Due to the non-concavity of the objective function in (\ref{equa:P0:1}), (P1) is generally difficult to be solved optimally.

In the following, we introduce two conventional energy beamforming designs and show the effect of the non-linear EH model in (\ref{equa:non-linear}) on the performance of the harvested DC energy.

\subsection{Multi-Beam Energy Transmission} \label{sec:MB}
In this design, the energy covariance matrix ${\mv S(t)}$ remains unchanged over the block, i.e., $\mv S(t)=\mv S, \forall t\in \mathcal T$. In this case, the min-DC-energy maximization problem in (P1) is reformulated as
\begin{align}
{\mathtt{(P2)}}: \max\limits_{\bm{S}}~&\min\limits_{k\in\mathcal K}~~Q^{\rm DC}_k(\mv h^H_k\mv S\mv h_k)\label{equa:multi_beam:1} \\
{\mathrm{s.t.}} ~&{\rm{tr}}({\mv{S}})\le P_{\rm max} \label{equa:max_power}\\
&\mv{S} \succeq 0. \label{equa:tr_S}
\end{align}
It is observed that function $Q_k^{\rm DC}(\mv h_k^H \mv S \mv h_k)$ in (\ref{equa:multi_beam:1}) is non-decreasing with respect to $\mv h^H_k\mv S\mv h_k$. Accordingly, after some manipulation and by introducing an auxiliary variable $E$, problem (P2) can be recast as
\begin{align}
{\mathtt{(P2.1)}}: \max\limits_{\bm{S}, E}&~~E \label{equa:p1.1} \\
{\mathrm{s.t.}}&~~\mv h^H_k\mv S\mv h_k\ge E,\ \forall k\in\mathcal K \label{equa:p1.1:E}\\
&~~(\ref{equa:max_power})~\text{and}~(\ref{equa:tr_S}). \nonumber
\end{align}
Problem (P2.1) is a semi-definite program (SDP) \cite{M.Grant_And_S.Boyd}, which can be efficiently solved using standard convex optimization tools such as CVX \cite{cvx}. Let $\mv S^{\star}$ denote the optimal solution to (P2), then the corresponding minimum harvested DC energy among all the ERs is $T\min\{Q^{\rm DC}_1(\mv S^{\star}),\ldots,Q^{\rm DC}_K(\mv S^{\star})\}$.

It is worth emphasizing that in general, the optimal energy covariance matrix $\mv S^{\star}$ may be of high rank, such that more than one energy beams are required to balance the harvested DC energy among all the ERs for fairness. This is the case especially when the number of ERs $K$ becomes sufficiently large \cite{XuWCL}. Furthermore, it can be verified that under the special case with the linear EH model, the multi-beam energy transmission is optimal for maximizing the harvested DC energy among the ERs. As such, the multi-beam energy transmission has been widely adopted in the literature \cite{XuWCL,XuSWIPT,LiuSWIPT}.


\subsection{TDMA-Based Energy Transmission}
In the TDMA-based design, the time block of interest $\mathcal T$ is divided into $K$ time slots with optimizable durations denoted as $\{\tau_n\}_{n=1}^K$, where $\tau_n \ge 0, \forall n\in \mathcal K$, and $\sum_{n\in\mathcal K}\tau_n=T$. In each slot $n\in \mathcal K$, the ET designs the energy beamforming matrices for maximizing the harvested DC energy at a particular ER $n$ with $\mv S_n^\star = P_{\rm max} \frac{\bm h_n \bm h_n^H}{|\bm h_n|^2}$. Accordingly, the harvested DC power for each ER $k\in \mathcal K$ at each slot $n$ is given as $Q_k^{\rm DC}(\mv h_k^H \mv S_n^\star \mv h_k)$. In this case, we only need to optimize the time allocation $\{\tau_n\}$ to maximize the minimum harvest DC energy among all the ERs, for which the optimization problem is expressed as
\begin{align}
{\mathtt{(P4)}}:\max\limits_{\{\tau_n\ge 0\}}~&\min\limits_{k\in\mathcal K}~\sum\limits_{n\in\mathcal K}\tau_n Q^{\rm DC}_{k}({\mv h_k^H\mv S^{\star}_n\mv h_k})\label{equa:3:ob} \\
{\mathrm{s.t.}}~~~& \sum\limits_{n\in\mathcal K}\tau_n=T. \label{equa:T}
\end{align}
By introducing an auxiliary variable $E$, problem (P4) is re-expressed as the following linear program (LP) \cite{M.Grant_And_S.Boyd}:
\begin{align}
{\mathtt{(P4.1)}}:\max\limits_{\{\tau_n\ge0\}, E}~&~E \label{equa:3.1:ob} \\
{\mathrm{s.t.}}~~~~&\sum\limits_{n\in\mathcal K}\tau_n Q^{\rm DC}_{k}(\mv S^{\star}_n)\ge E,~\forall k\in\mathcal K \label{equa:4.5:Eob}\\
& (\ref{equa:T}),\nonumber
\end{align}
which can be optimally solved by CVX.

Note that under the special case of linear EH models, the TDMA-based design is a suboptimal solution to the min-DC-energy maximization problem for multiuser WPT, and thus performs inferior to the optimal multi-beam energy transmission in Section \ref{sec:MB}. Nevertheless, this may not hold in general under the practical non-linear EH model of our interest, as will be illustrated in the following simple example.
\begin{example}\label{Example:1}
Consider the case with $K=2$ ERs, in which the number of transmit antennas at the ET is $M=4$, and the maximum transmit power is $P_{\rm max}=15$ W. Suppose that the channel vectors from the ET to the two ERs are orthogonal, given by $\mv h_1=\{10^{-4},10^{-4},0,0\}^T$ and $\mv h_2=\{0,0,10^{-4},10^{-4}\}^T$, respectively. As for the non-linear EH model, we set $Q^{\rm max} = 10.73$ mW, $a=0.2308$, and $b=5.365$ \cite{Clerckx}. Under this setup and by considering a unit time duration with $T=1$, it can be shown that for the multi-beam energy transmission, the achieved RF power at each ER is $1.5$ mW and the corresponding average DC power at each ER is $0.9127$ mW. By contrast, for the TDMA-based energy transmission, the achieved RF power by ET 1 at the two slots is $3$ mW and $0$ mW, and the corresponding DC power is $1.9666$ mW and $0$ mW, respectively; while the achieved RF power by ET 2 at the two slots is $0$ mW and $3$ mW, and the corresponding DC power is $0$ mW and $1.9666$ mW, respectively. Accordingly, with an optimal equal time allocation between the two slots, the average DC power at each ER is $0.9833$ mW, which is higher than that achieved by the multi-beam energy transmission. This example clearly shows that the TDMA-based design outperforms the multi-beam energy transmission, which is due to that for the multi-beam design, the transmit power is allocated to two orthogonal beamforming vectors, thus leading to a small received RF power at each ER that falls in the convex region for RF-to-DC conversion. In this regime, we can increase the RF-to-DC power conversion efficiency by increasing the input RF power. It is thus beneficial to use the TDMA-based energy transmission for achieving larger received RF power at each slot to improve the overall DC power performance. Based on this observation, we are motivated to propose a new energy beamforming design to exploit both benefits of multi-beam and TDMA-based energy transmission under the non-linear EH model.
\end{example}

\section{Time-Division Energy Beamforming}
In this section, we propose a new time-division energy beamforming approach. In this design, the transmission block $\mathcal T$ is divided into $K$ slots, each with optimizable slot $\tau_n$. In each slot $n \in \mathcal K$, the transmit energy covariance is $\mv S_n$, which is generally of high rank and optimizable. Accordingly, the received RF power at ER $k$ in slot $n$ is $Q_k^{\rm RF}=\mv h^H_k\mv S_n\mv h_k$, and the harvested DC power is $Q^{\rm DC}_{k}({\mv h}^{H}_k\mv S_n{\mv h}_k)$. Hence, the average harvested DC energy at each ER $k$ is given by $\sum_{n\in\mathcal K}\tau_n Q^{\rm DC}_{k}({\mv h}^{H}_k\mv S_n{\mv h}_k)$. Under the proposed time-division energy beamforming design, the min-energy maximization problem is formulated as
\begin{align}
{\mathtt{(P5)}}:\max\limits_{\{\tau_n\ge0\},\{\bm S_n\}}~&\min\limits_{k\in\mathcal K}\sum\limits_{n\in\mathcal K}\tau_n Q^{\rm DC}_{k}({\mv h}^{H}_k\mv S_n{\mv h}_k)\label{equa:3T:ob} \\
{\mathrm{s.t.}}~~~~~& {\rm{tr}}({\mv S_n})\le P_{\rm max},~\forall n\in\mathcal K\label{equa:3T:max}\\
& \mv S_n \succeq 0,~\forall n\in\mathcal K \label{equa:3T:S}\\
&(\ref{equa:T}).\nonumber
\end{align}

Note that for the special case of $\mv S_n = \mv S$, $n\in\mathcal K$, our proposed design reduces to the conventional multi-beam energy transmission; while if $\mv S_n = P_{\rm max} \frac{\bm h_n \bm h_n^H}{|\bm h_n|^2}$, it becomes the TDMA-based design. Therefore, our proposed design can exploit both benefits of the two conventional designs based on the non-linear EH model.

Problem (P5) is non-convex due to the coupling between the time durations $\{\tau_n\}$ and covariance matrices $\{\mv S_n\}$, and thus is very challenging to be optimally solved. To resolve this issue, we propose an efficient alternating-optimization-based algorithm to solve (P5) by optimizing $\{\tau_n\}$ and $\{\mv S_n\}$ in an alternating manner \cite{OALiu}.

\subsubsection{Optimization of $\{\mv S_n\}$ under Given $\{\tau_n\}$}
Under given time durations $\{\tau_n\}$, the optimization of $\{\mv S_n\}$ can be expressed as
\begin{align}
{\mathtt{(P5.1)}}: \max\limits_{\{\bm S_n\}}~&\min\limits_{k\in\mathcal K}~\sum\limits_{n\in\mathcal K}\tau_n Q^{\rm DC}_{k}({\bm h}^{H}_k\bm S_n{\bm h}_k)\label{equa:4.1:o} \\
{\mathrm{s.t.}}~~&(\ref{equa:3T:max})~\text{and}~(\ref{equa:3T:S}).
\nonumber
\end{align}
The above problem (P5.1) is still non-convex due to the non-convex objective function. To solve this problem, we propose to update the beamforming matrices $\{\mv S_n\}$ by applying the SCA technique \cite{SCALiu,SCALuo}. Consider the (inner) iteration $l\ge 1$, in which the current point of $\{\mv S_n\}$ is $\{\mv S^{(l-1)}_n\}$. At the current point $\{\mv S^{(l-1)}_n\}$, we approximate the non-convex objective function of (\ref{equa:4.1:o}) by its first-order Taylor expansion, given by
\begin{align}
&Q^{\rm DC}_{k}({\bm h}^{H}_k\bm S_n{\bm h}_k)\approx \nonumber\\
& Q^{\rm DC}_{k}({\bm h}^{H}_k\bm S^{(l-1)}_n{\bm h}_k)+\nabla{Q^{\rm DC}_{k}}({\bm h}^{H}_k\bm S^{(l-1)}_n{\bm h}_k){\bm h}^{H}_k\bm S_n{\bm h}_k,
\end{align}
where
\begin{align*}
\nabla{Q^{\rm DC}_{k}}({\mv h}^{H}_k\bm S^{(l-1)}_n{\bm h}_k)=\frac{Q^{\rm max}ae^{-a({\bm h}^{H}_k\bm S^{(l-1)}_n{\bm h}_k)}}{(1+e^{-a({\bm h}^{H}_k\bm S^{(l-1)}_n{\bm h}_k-b)})^2(1-\Omega)}.
\end{align*}
Since $Q^{\rm DC}_{k}({\bm h}^{H}_k\bm S_n{\bm h}_k)$ is not convex nor concave, we need to define a trust region to ensure the approximation accuracy \cite{Trust}. The following trust region constraints are imposed:
\begin{align}
|{\bm h}^{H}_k\bm S_n{\bm h}_k-{\bm h}^{H}_k\bm S^{(l-1)}_n{\bm h}_k|\le \Gamma_t, \forall k\in\mathcal K,\ \forall n\in\mathcal K, \label{equa:trust_region}
\end{align}
in which $\Gamma_t>0$ denotes the radius of the trust region.

As a result, problem (P5.1) is approximated as
\begin{align}
{\mathtt{(P5.2)}}:\max\limits_{\{\bm S_n\}}~&\min\limits_{k\in\mathcal K}\sum\limits_{n\in\mathcal K}\tau_n \nabla{Q^{\rm DC}_{k}}({\bm h}^{H}_k\bm S^{(l-1)}_n{\bm h}_k){\bm h}^{H}_k\bm S_n{\bm h}_k\nonumber \\
{\mathrm{s.t.}}~~
&(\ref{equa:3T:max}),~(\ref{equa:3T:S}),~\text{and}\ (\ref{equa:trust_region}).\nonumber
\end{align}
By introducing an auxiliary variable $E$, problem (P5.2) is equivalent to
\begin{align}
&{\mathtt{(P5.3)}}:\nonumber\\
&\max\limits_{\{\bm S_n\}, E}~E\label{equa:4.3:E} \\
&~~~{\mathrm{s.t.}}~~\sum\limits_{n\in\mathcal K}\tau_n \nabla{Q^{\rm DC}_{k}}({\bm h}^{H}_k\bm S^{(-1)}_n{\bm h}_k){\bm h}^{H}_k\bm S_n{\bm h}_k\ge E,~\forall k\in\mathcal K \label{equa:4.3:Eob}\\
&~~~~~~~~~(\ref{equa:3T:max}),~(\ref{equa:3T:S}),~\text{and}~(\ref{equa:trust_region}) .\nonumber
\end{align}
Problem (P5.3) is convex and thus can be solved by e.g., CVX. Let $\{\mv S^{(l)}_n\}$ denote the optimal solution to (P5.3) at iteration $l$. By taking $\{\mv S^{(l)}_n\}$ into (\ref{equa:4.1:o}), if the objective value increases, then we replace the current point by $\{\mv S^{(l)}_n\}$ and go to the next iteration; otherwise, we reduce $\Gamma_l$ and go back to solve problem (P5.3) until $\Gamma_l$ is less than the tolerance $\varepsilon$, i.e., $\Gamma_l\le\varepsilon$.

\subsubsection{Optimization of $\{\tau_n\}$ under Given $\{\mv S_n\}$}
We optimize the time duration $\{\tau_n\}$ under given beamforming matrices $\{\mv S_n\}$. In this case, problem (P4) can be expressed as
\begin{align}
{\mathtt{(P5.4)}}:\max\limits_{\{\tau_n\ge0\}}~&\min\limits_{k\in\mathcal K}~\sum\limits_{n\in\mathcal K}\tau_n Q^{\rm DC}_{k}({\mv h}^{H}_k\mv S_n{\mv h}_k)\label{equa:4.4:ob} \\
{\mathrm{s.t.}}~~~& (\ref{equa:T}).\nonumber
\end{align}
Problem (P5.4) is similar to problem (P4), and thus can be transformed into an LP and solved by e.g., CVX.

By solving problem (P5.1) and (P5.4) in an alternating manner, problem (P5) can be finally solved. The detailed algorithm is summarized as Algorithm 1.
\begin{algorithm}[t]
\caption{for Solving Problem (P5)}
\begin{itemize}
\item[1:] {\bf Initialize} the termination tolerance $\epsilon>0$, the trust region radius bound $\varepsilon>0$, the outer and inner iteration indices $p=1$ and $l=1$, the initial time duration $\tau^{(p-1)}_n=1/K, \forall n \in \mathcal K$, and initial covariance matrices $\{\mv S^{(p-1)}_n\}$. Choose $\Gamma_l>0$.
\item[2:] {\bf Repeat:}
\item[3:] \ \ \ Solve problem (P5.1) to find the optimal beamforming matrices $\{\mv S^{(p)}_n\}$ under given $\{\tau^{(p-1)}_n\}$:
\item[4:]\ \ \ {\bf Repeat:}
\item[5:]\ \ \ \ Solve (P5.3) under given $\{\tau^{(p-1)}_n\}$ and $\{\mv S^{(p-1)}_n\}$ to obtain solution $\{\mv S^{(l)}_n\}$. {\bf If} the objective value of (\ref{equa:4.1:o}) increases, then update $\mv {\hat S}^{(p)}_n=\mv S^{(l)}_n$ and set $l=l+1$; {\bf otherwise}, set $\Gamma_l=\Gamma_l/2$.
\item[6:]\ \ \ {\bf Until} $\Gamma_l\le\varepsilon$, {\bf return} $\{\mv S^{(p)}_n\}=\{\mv {\hat S}^{(p)}_n\}$ and $l=0$.
\item[7:]\ \ \ Solve problem (P5.4) with the obtained $\{\mv S^{(p)}_n\}$ to find the optimal $\{\tau^{(p)}_n\}$ and set $p=p+1$.
\item[8:]{\bf Until} the increase of the objective function in (P5) is smaller than $\epsilon$.
\end{itemize}
\end{algorithm}

\section{Numerical Results}\label{sec:NR}
In this section, we provide numerical results to validate the performance of our proposed time-division energy beamforming design, as compared to the multi-beam and TDMA-based energy transmission designs. Furthermore, we also consider the isotropic beamforming design as another benchmark scheme, in which the transmit energy covariance matrix is set to be $\mv S_{\rm ist}=\frac{P_{\rm max}}{M}\mv I$ over the whole transmission block, such that the wireless energy is broadcast isotropically over space.


In the simulations, we consider the Rician fading channel from the ET to ER $k$, given by
\begin{align}
\mv H_k = \sqrt{\frac{K_R}{1+K_R}}\mv H^{\rm{LOS}}_k+\sqrt{\frac{1}{1+K_R}}\mv H^{\rm{NLOS}}_k. \label{equa:channels}
\end{align}
Here, the Rician factor is set to be $K_R=5$ dB, $\mv H^{\rm LOS}_k$ denotes the LOS component, and $\mv H^{\rm NLOS}_k$ denotes the non-LOS Rayleigh fading component. We consider the uniform linear antenna array model at the ET, i.e., each row of $\mv H^{\rm LOS}_k$ is $\sqrt{g}[1\ e^{i\theta_k}\ ...\ e^{i(M-1)\theta_k}]$ with $\theta_k=-\frac{2\pi\kappa\sin(\phi_k)}{\lambda}$, in which $\kappa=\frac{\lambda}{2}$ denotes the spacing between two successive antenna elements at the ET, $\lambda$ is the carrier wavelength, and $\phi_k=-\frac{5}{12}\pi+\frac{2}{K}\pi(k-1)$ is the direction of the $k$-th ER from the ET. Furthermore, $g = \zeta_0d^{-3}$ denotes the average channel power gain, in which $d = 4$ meters (m) is the distance between the ET and each ER and $\zeta_0=-30$~dB denotes the channel power gain at a reference distance of $d_0 = 1$ m. Furthermore, we set the transmit antenna gain at the ET as $10$ dBi and the receive antenna gain at each ER as $2.8$ dBi. The number of ERs is set to be $K=30$. As for the non-linear EH model, the parameters are same as those in Example \ref{Example:1}.

\begin{figure}
\centering
 \epsfxsize=1\linewidth
    \includegraphics[width=6cm]{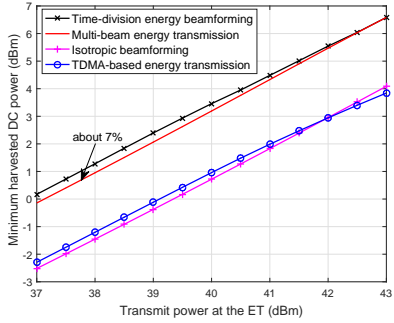}
\caption{The minimum harvested DC power among the ERs versus the transmit power at the ET $P_{\rm max}$ with $M=4$.} \label{fig:power}
\vspace{-1em}
\end{figure}

\begin{figure}
\centering
 \epsfxsize=1\linewidth
    \includegraphics[width=6cm]{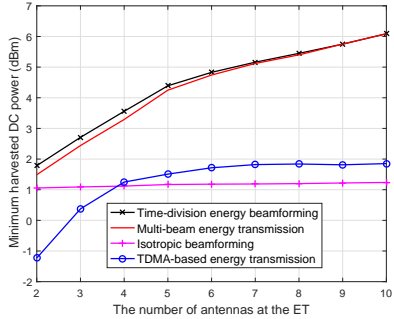}
\caption{The minimum harvested DC power among all ERs versus the number of transmit antennas $M$ with $P_{\rm max}=40$ dBm.} \label{fig:antenna}
\vspace{-2em}
\end{figure}
Fig. \ref{fig:power} shows the average minimum harvested DC power among the ERs versus the transmit power at the ET $P_{\rm max}$, where the transmit antenna is set to be $M=4$. It is observed that our proposed time-division energy beamforming design outperforms all the other three conventional designs. More specifically, when the transmit power is $P_{\rm max} = 38$ dBm, our proposed design achieves around 7$\%$ higher harvested DC energy than the multi-beam energy transmission, as the received RF power falls in the convex region for RF-to-DC power conversion, which validates the effectiveness of the proposed design. When the transmit power becomes large (e.g., $P_{\rm max} \ge 42$ dBm), our proposed design and multi-beam energy transmission have the same performance. This is due to the fact that the received RF power is in the concave region in this case, and thus the proposed time-division energy beamforming reduces to the conventional multi-beam energy transmission that is optimal.

Fig. \ref{fig:antenna} shows the minimum harvested DC power among ERs versus the number of antennas at the ET $M$ with the transmit power being $P_{\rm max}=40$ dBm. It is observed that when $2\le M\le 6$, our proposed design outperforms the multi-beam energy transmission. When the number of antennas becomes large, these two designs have the same performance. This is because with more transmit antennas, higher array gains can be exploited to increase the received RF power, such that the RF-to-DC conversion may work in the concave regime. Therefore, the proposed design achieves the same performance as the multi-beam energy transmission. Note that for the isotropic beamforming design, the resultant harvested DC energy is observed to almost keep uncharged. This is due to that in the isotropic beamforming design, the energy is broadcasted isotropically no matter how many antennas are employed. \vspace{-0.5em}

\section{Conclusion}
This letter investigated the multiuser MISO WPT system under the non-linear EH model. Under this setup, we found that the conventional multi-beam energy transmission, which is optimal for min-DC-energy maximization under the linear EH models, are not optimal any longer under the practical non-linear EH models. To overcome this drawback, we proposed a novel time-division energy beamforming, in which different energy beamforming matrices are time shared to exploit the non-linear nature of RF-to-DC energy conversion. We jointly optimized the beamforming matrices and the corresponding transmission time duration to maximize the minimum harvested DC energy among all the ERs. Numerical results showed that our proposed design indeed improved the performance especially when the received RF power falls in the convex region for RF-to-DC conversion.

%

\end{document}